\documentclass[
showpacs,
aps,
prl,
twocolumn,
superscriptaddress,
tightenlines
]{revtex4-1}

\usepackage{times}
\usepackage{graphicx}
\usepackage{amsmath}
\usepackage{amssymb}
\def\vd{v_d}

\def\nph{n^{\mbox{\scriptsize ph}}}

\def\tbg{T_{BG}}
\def\te{T_e}
\def\nim{\nu_{im}}
\def\nph{\nu_{ph}}

\newcommand{\be}{\begin{equation}}
\newcommand{\ee}{\end{equation}}

\newcommand{\req}[1]{Eq.\,(\ref{#1})}

\newcommand{\rfig}[1]{Fig.\,\ref{#1}}

\begin{document}
\title{
Bloch-Gr\"{u}neisen nonlinearity of electron transport in GaAs/AlGaAs heterostructures
}
\author{O.\,E. Raichev}
\affiliation{Institute of Semiconductor Physics, National Academy of Sciences of Ukraine, Prospekt Nauki 41, 03028 Kiev, Ukraine}
\author{A.\,T. Hatke}
\email[Present address: ]{Department of Physics and Astronomy, Purdue University, West Lafayette, Indiana 47907, USA.}
\affiliation{National High Magnetic Field Laboratory, Tallahassee, Florida 32310, USA}
\author{M.\,A. Zudov}
\email[Corresponding author: ]{zudov@physics.umn.edu}
\affiliation{School of Physics and Astronomy, University of Minnesota, Minneapolis, Minnesota 55455, USA}
\author{J.\,L. Reno}
\affiliation{CINT, Sandia National Laboratories, Albuquerque, New Mexico 87185, USA}
\received{\today}

\begin{abstract}
We report on nonlinear transport measurements in a two-dimensional electron gas hosted in a GaAs/AlGaAs heterostructure. 
Upon application of direct current, the low-temperature differential resistivity acquires a positive correction, which exhibits a pronounced maximum followed by a plateau.
With increasing temperature, the nonlinearity diminishes and disappears.
These observations can be understood in terms of a crossover from the Bloch-Gr\"uneisen regime to the quasielastic scattering regime as the electrons are heated by the current. 
Calculations considering the interaction of electrons with acoustic phonons provide a reasonable description of our experimental findings. 
\end{abstract}
\received{26 June 2017} 
\maketitle

Nonlinear transport in semiconductors \citep{ting:1992}, characterized by significant changes in the resistance takes place at strong electric fields (drift velocities $\vd \gtrsim 10^7$ cm/s) when electrons gain enough energy to cause intense optical-phonon emission or intervalley transfer. 
Two-dimensional electron gases (2DEGs) in heterostructures usually require similar conditions to show nonlinear behavior \citep{ting:1992,danilchenko:2004,sun:2007}. 
However, in a 2DEG placed in a magnetic field, nonlinear effects are prominent even at moderate $\vd$ as a result of Landau quantization. 
One such effect is Hall field-induced resistance oscillations \cite{yang:2002,bykov:2005c,zhang:2007a,bykov:2007,zhang:2008,hatke:2009c,hatke:2011a,shi:2014b,shi:2017b} which originate from electron transitions between Landau levels, tilted by the Hall field, owing to electron backscattering off impurities \citep{yang:2002,vavilov:2007,lei:2007}.
When the current density $j$ is reduced (or the magnetic field $B$ is increased) such transitions are no longer possible and the differential resistivity is \emph{suppressed} \citep{zhang:2007a,zhang:2007b,vavilov:2007,lei:2007,hatke:2010a,hatke:2012d} (and can even vanish \citep{bykov:2007,hatke:2010a}).
%This suppression originates from the dc field-induced correction to the electron distribution function and direct suppression of electron-disorder scattering \citep{vavilov:2007,lei:2007}.
Both of the above nonlinearities disappear as the magnetic field is lowered due to increasing overlap between the Landau levels.

In this Rapid Communication we report on another kind of nonlinearity which takes place at \emph{zero magnetic field} and is characterized by an \emph{increase} in the resistance at \emph{moderate} drift velocities ($\vd \lesssim 10^6$ cm/s). 
This nonlinearity originates from a crossover between two distinct regimes of electron-phonon interaction, owing to the existence of the maximal energy transferred in the process of phonon emission.
The temperature corresponding to this energy is known as the Bloch-Gr\"uneisen (BG) temperature \citep{bloch:1930,gruneisen:1933}, $\tbg=2 s p_F/k_B$, where $s$ is the sound velocity and $p_F$ is the electron Fermi momentum. 
If the electron temperature $\te < \tbg$, electron-phonon scattering is suppressed, as only phonons with energies smaller than $k_B T_e$ can be emitted or absorbed.
If $\te > \tbg$, there are no such restrictions, so the scattering no longer depends on the electron distribution and becomes effectively elastic \citep{note:0}. 

The importance of the BG regime ($\te < \tbg$) is recognized when electron-phonon scattering is of key significance, e.g, in the energy relaxation of nonequilibrium electrons \citep{barlow:1988,ma:1991} or in phonon-drag thermoelectricity \citep{ruf:1988}.
It is harder to detect the BG regime in the resistance measurements because it takes place at low temperatures ($\tbg \lesssim 10$ K in a typical 2DEG) when the resistance is limited by impurity scattering. 
Observations of the BG regime in the temperature dependence of the resistance have been accomplished in high-mobility GaAs/AlGaAs heterostructures \citep{stormer:1990} and in graphene layers \citep{efetov:2010} with high electron density. 
As we demonstrate below, the nonlinear response in a 2DEG provides an easier and a more efficient way to detect the BG regime; 
with increasing current, the 2DEG is heated and undergoes a transition from the BG regime to the quasielastic scattering regime, manifested by a step-like increase in the resistance.
    
%\section{Experiment}
While we have investigated several samples with a similar outcome, here we focus on data from a Hall bar (width $w=25\,\mu$m) fabricated from a GaAs/AlGaAs heterostructure (EA0761) with density $n_e \approx 1.6 \times 10^{11}$ cm$^{-2}$ and mobility $\mu \approx 4.6 \times 10^{6}$ cm$^{2}/$V\,s.
The differential resistance was recorded using a low-frequency lock-in technique as a function of direct current $I$ at various coolant temperatures $T$ from 2 to 12 K.

%%%%%%%%%%%%%%%%%%%%%%%%%%%%%%%%%%%%%%%%%%%%%%%%%
%fig 2
\begin{figure}[t]
\includegraphics{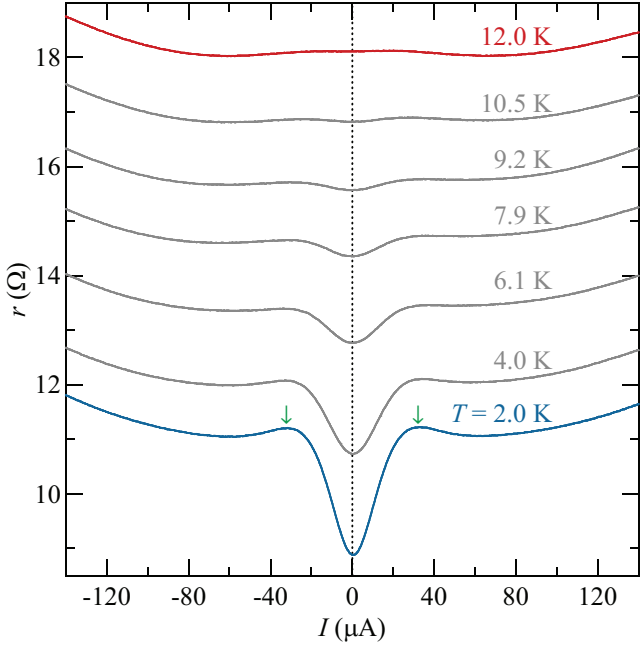}
\vspace{-0.1 in}
\caption{(Color online) 
$r$ vs $I$ measured at $T$ from $2.0$ to $12.0$ K, as marked. 
%The traces are \emph{not} offset.
Arrows mark the crossover from BG to the quasielastic regime.
}
\vspace{-0.1 in}
\label{fig1}
\end{figure}
%%%%%%%%%%%%%%%%%%%%%%%%%%%%%%%%%%%%%%%%%%%%%%%%%
In \rfig{fig1} we present the differential resistivity $r$ as a function of current $I$ measured at various $T$, as marked.
At the low $T$, $r$ first rapidly increases with $I$, exhibits a maximum (cf. $\downarrow$), and then a slight decrease, followed by subsequent growth at higher $I$. 
When $T$ is raised, $r$ increases at all $I$, and, concurrently, the nonlinearity observed at $I \lesssim 40$ $\mu$A gradually diminishes and eventually disappears.
The increase of $r$ at $I \gtrsim 80$ $\mu$A, however, remains essentially unchanged at all $T$ studied. 
The contrasting $T$ dependencies of lower-$I$ and higher-$I$ nonlinearities suggest that these are characterized by different energy scales and thus are of distinct physical origins.

\begin{figure}[b]
\vspace{-0.10 in}
\includegraphics{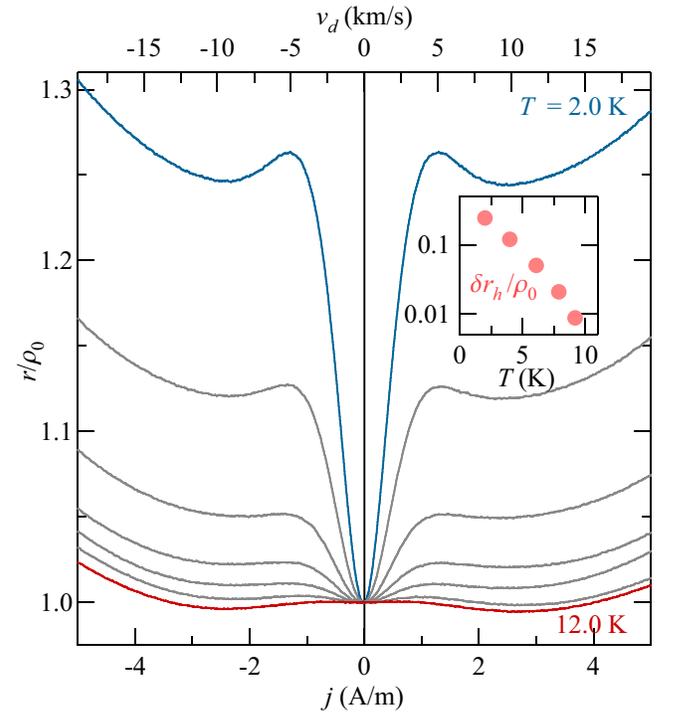}
\vspace{-0.1 in}
\caption{(Color online) 
$r/\rho_0$ vs $j$ (bottom axis) and $\vd$ (top axis) at the same $T$ as in \rfig{fig1}.
Inset: $\delta r_h/\rho_0$ vs $T$.
}
%\vspace{-0.1 in}
\label{fig2}
\end{figure}

To examine our findings in more detail, we normalize $r$ by its linear-response value $\rho_0$ at each $T$ and present the results in \rfig{fig2} as a function of $j = I/w$ (bottom axis) and electron drift velocity $\vd = j/en_e$ (top axis).
To quantify the lower-$I$ nonlinearity, we introduce $\delta r_h  = r_h - \rho_0$, where $r_h$ is the value of the differential resistivity at the broad minimum occurring near $\vd = 10$ km/s.
In the inset, we present the $T$ dependence of $\delta r_h/\rho_0$, which highlights its rapid disappearance with increasing $T$.

We next argue that our observations can be explained in terms of the modification of electron-phonon scattering rate $\nph$ due to $I$-induced heating of the 2DEG. 
A considerable $T$ dependence of the resistivity $\rho \simeq m^\star(\nim+\nph)/e^2 n_e$ ($m^\star$ is the effective mass) suggests that $\nph$ is comparable to the electron-impurity scattering rate $\nim$. 
The nature of the dependence of $\nph$ on $T$ and on the electron temperature $\te$ is different;  increasing $T$ leads to a steady growth of $\nph$ because of an increasing number of phonons, while increasing $\te$ enhances $\nph$ only in the BG regime, $T < \te < \tbg$ ($\tbg \approx 7$ K in our sample). 
For stronger heating, $\te > \tbg$, one may expect a weaker and, generally, nonmonotonic change of $\rho$ when $k_B \te$ becomes comparable to the chemical potential $\eta$. 

%%%%%%%%%%%%%%%%%%%%%%%%%%%%%%%%%%%%%%%%%%%%%%%%%
%fig 2
%%%%%%%%%%%%%%%%%%%%%%%%%%%%%%%%%%%%%%%%%%%%%%%%%
 
%\section{Theory and discussion} 
To demonstrate that the mechanisms of the nonlinearity discussed above are indeed relevant in our experiment, we have carried out detailed calculations of the differential resistivity. 
The distribution function of electrons $f_{{\bf p}}$ is found from the classical Boltzmann equation ${\bf v}_{\bf p} \cdot 
\nabla f_{{\bf p}} + e{\bf E} \cdot (\partial f_{{\bf p}}/\partial {\bf p}) =J_{{\bf p}}$, where ${\bf v}_{\bf p}={\bf p}/m^\star$ is the electron velocity and ${\bf E}$ is the electric field. 
The collision integral $J_{{\bf p}}$ comprises electron-impurity, electron-phonon, and electron-electron contributions. 
Assuming that the isotropic part of $f_{{\bf p}}$ is controlled by electron-electron collisions, we write it in the Fermi-like form:
\begin{equation}
f_{\varepsilon}=\left[ \exp \frac{\varepsilon -\eta}{\te} + 1 \right]^{-1}\,,
\label{eq.f}
\end{equation}
where $\varepsilon=\varepsilon_p=p^2/2m^\star$ is the electron energy (here and below we set $k_B$ to unity). 
The remaining anisotropic part of $f_{{\bf p}}$ is determined by electron-impurity and electron-phonon scattering. 
For moderate $I$ relevant to our experiment, this part is small compared to the isotropic one and can be found by linearizing the kinetic equation. 
In spite of the inelastic nature of the electron-phonon scattering, the solution of the linearized equation can be represented analytically, owing to the smallness of phonon energies compared to the average electron energy,
\begin{equation}
f_{{\bf p}} \simeq f_{\varepsilon} - \tau(\varepsilon){\bf v}_{\bf p} \cdot \left[ e {\bf E}  
\frac{\partial f_{\varepsilon} }{\partial \varepsilon} + \nabla f_{\varepsilon} \right]\,.
\end{equation}
Here, $\tau(\varepsilon)=1/[\nim(\varepsilon)+\nph(\varepsilon)]$ and 
\begin{eqnarray}
\nph(\varepsilon) = \frac{2 m^\star}{\hbar^3} \sum_{i} \int_{0}^{\pi} \frac{d \theta}{\pi} 
(1-\cos \theta)  \int_0^{\infty} \frac{d q_z}{\pi}
\nonumber \\
\times  \int_{0}^{\pi} \frac{d \varphi_q}{\pi} 
C_{i {\bf Q}} I(q_z) \left[N_{\omega}-N^{e}_{\omega} + N^{e}_{\omega}(1+N^{e}_{\omega}) \frac{\hbar \omega}{\te}\right],
\label{eq.th.1}
%3
\end{eqnarray}
where $i$ labels phonon modes, $\omega=\omega_{i\bf Q}$ is the phonon frequency, ${\bf Q}=({\bf q},q_z)$ is the phonon wave vector, ${\bf q}$ is its in-plane component described by magnitude $q=2 k_{\varepsilon}\sin(\theta/2)$ ($k_{\varepsilon}=\sqrt{2m^\star \varepsilon}/\hbar$) and polar angle $\varphi_q$, $\theta$ is the scattering angle, $N_{\omega}=[e^{\hbar \omega/T}-1]^{-1}$ ($N^{e}_{\omega}=1/[e^{\hbar \omega/\te}-1]$) is the Planck distribution with $T$ ($\te$), $C_{i {\bf Q}}$ is the squared matrix element of the electron-phonon interaction, and $I(q_z)=|\int dz |\psi(z)|^2 e^{i q_z z}|^2$ is the squared overlap determined by the electron wave function $\psi(z)$. 

It is necessary to take into account the spatial dependence of $f_{{\bf p}}$ because heating of the 2DEG appears to be inhomogeneous due to the heat transfer caused by drift and diffusion in a finite-size sample. 
We have found that this spatial dependence leads to a sizable modification of $r$ compared to a homogeneous approximation and is likely a reason for slight asymmetry with respect to $I$ direction observed on Figs. 1 and 2. 
The approximation in Eq. (2) means that $f_{{\bf p}}$ depends on coordinate ${\bf r}$ parametrically, through $\eta$ and $\te$. 
As $\eta$ is expressed through $\te$ by the requirement of fixed local density, it suffices to find $\te({\bf r})$, for which we use the energy balance equation 
\begin{equation}
\nabla \cdot {\bf G}-{\bf j} \cdot {\bf E} + P =0\,, 
%4 
\end{equation}
where $P=-2 (2 \pi \hbar)^{-2} \int d {\bf p} J_{\bf p} \varepsilon_p$ is the energy density dissipated per unit time due to collisions and ${\bf G}= 2 (2 \pi \hbar)^{-2} \int d {\bf p} {\bf v}_{\bf p} \varepsilon_p f_{{\bf p}}$ is the energy density flux. 
Application of Eq. (2) leads to the standard expressions \citep{note:vasko}
\begin{eqnarray}
{\bf j}=\sigma [{\bf E} - \nabla \eta/e - S \nabla \te]\,, \\
{\bf G}= (S \te + \eta/e){\bf j} -\kappa \nabla \te\,,
\end{eqnarray}
where $S$ is the Seebeck coefficient, $\kappa$ is the electronic thermal conductivity, and $\sigma=e^2 n_e \tau/m^\star$ is the electrical conductivity, where $\tau$ is found using a standard averaging procedure $\tau = (m^\star/\pi \hbar^2 n_e) \int_0^{\infty} d \varepsilon \left(-\partial f_{\varepsilon}/\partial \varepsilon \right) \varepsilon \tau (\varepsilon)$. 
We neglect the energy dependence of $\nim(\varepsilon)$ because it is determined by the scattering potential which is generally not known, so the dependence $\tau(\varepsilon)$ comes from $\nph(\varepsilon)$ given by Eq. (3).  

Substituting Eqs. (5) and (6) into Eq. (4), one gets a nonlinear differential equation 
\begin{equation}
\nabla ( \kappa \nabla \te) - \te\, {\bf j} \cdot {\nabla S} + j^2/\sigma - P =0\,, 
%7 
\end{equation}
which has been solved numerically to find $\te({\bf r})$. 
Given our sample geometry, we assume that $\te$ depends only on the coordinate $x$ along the Hall bar. 
The differential resistance $dV/dI$ is found from $V=R(I)I$, where $R(I)=w^{-1} \int_{x_1}^{x_2} dx \rho(x)$ is the total resistance expressed through $\rho=1/\sigma$ (which depends on $I$ because of electron heating), and $x_1,x_2$ mark the locations of the voltage probes.
Since the voltage leads stay in equilibrium (regardless of $\te$), there is no thermoelectric contribution to the response.

Based on this formalism, we now present a qualitative analysis of the effects of electron heating on the resistance. 
Even if we assume that the 2DEG is degenerate [$\nph(\varepsilon) = \nph(\eta)$] and uniformly heated, its resistivity $\rho=m^\star(\nim+\nph)/e^2 n_e$ depends on $\te$ through $\nph$ given by Eq. (3). 
In particular, $\rho$ increases with $\te$ in the BG regime, $\te < \tbg$. 
When $\te \gg \tbg$, $N_\omega^e \approx \te/\hbar \omega-1/2$, the term in square brackets in \req{eq.th.1} reduces to $N_\omega + 1/2$, and $\nph$ (and hence $\rho$) becomes independent of $\te$. 
Therefore, with increasing $I$, the resistivity changes from $\rho_0$ to a saturated value $\rho_h$ (as $\rho_h$ is independent of $I$, $r$ also changes from $\rho_0$ to $r_h \equiv \rho_h$). 
The current at which this change takes place is determined by $\te \approx \tbg$ and corresponds to $\vd \approx 5$ km/s, according to our estimates. 

At higher $I$, such that $\te \sim \eta$, the degenerate approximation is no longer valid and there appears another nonlinearity associated with the energy dependence of $\nph(\varepsilon)$. 
The factors determining $\nph$ are the deformation and piezoelectric mechanisms of the electron-phonon interaction. 
For the first one, the rate increases with $\varepsilon$, while for the second one it decreases, so the resistance may depend on $I$ nonmonotonically when the Fermi distribution is broadened due to electron heating. 
When $\te \gtrsim \eta$, the average energy of the 2DEG starts to increase with $I$ and the deformation mechanism becomes more important, leading to the enhancement of resistance. 
Therefore, instead of a simple saturation at $r = r_h$, one expects a nonmonotonic dependence on $I$, with a steady growth at high $I$, in agreement with our data.

Further, we notice that thermoelectric effects cause the appearance of the term linear in $j$ in the balance equation (7). 
In nonsymmetric Hall bars, as the one used in our experiment, this term brings in the sensitivity of $\te({\bf r})$ to the direction of $I$.
Consequently, $r$ also becomes sensitive to the $I$ direction, and the asymmetry with respect to $j$ should increase at higher $I$ because the inhomogeneity increases with $\te$, owing to the enhancement of temperature gradients determining heat fluxes. 
The asymmetry in Figs. 1 and 2 very likely originates from the inhomogeneous heating described above.  

\begin{figure}[t]
\includegraphics{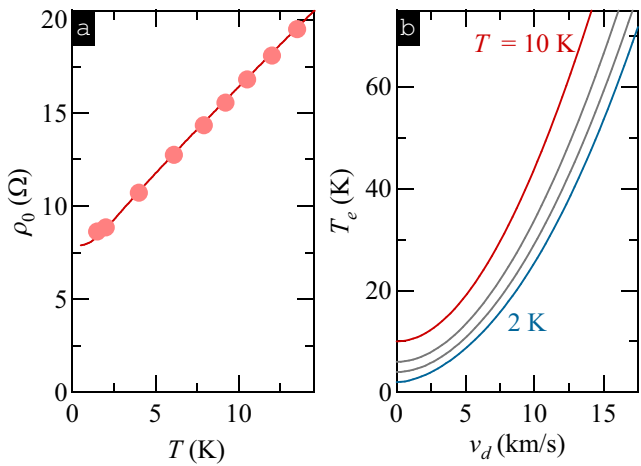}
\vspace{-0.1 in}
\caption{
(a) Measured (circles) and calculated (line) $\rho_0$ vs $T$.
(b) Calculated $\te$ vs $\vd$ for $T = 2,4,6$ and 10 K. 
}
\vspace{-0.1 in}
\label{fig3}  
\end{figure}

For a detailed numerical analysis, we use the model of isotropic phonons with $\omega_{i {\bf Q}}=s_i Q$ and the following expression for the squared matrix element of the electron-phonon interaction,
\begin{eqnarray}
C_{i {\bf Q}} = \frac{\hbar}{2 \rho_M s_i Q} \left[{\cal D}^2 Q^2 \delta_{i,l} + (eh_{14})^2 {\cal F}_i 
\right], \\
{\cal F}_l=\frac{9q^4q_z^2}{2Q^6}\left[1-\cos(4 \varphi_q)\right ], \nonumber \\ 
{\cal F}_t=\frac{2}{Q^6}
\left[q^2q_z^4+\frac{q^6}{8} + \left( q^4q_z^2-\frac{q^6}{8} \right) \cos(4 \varphi_q)\right]\,. \nonumber
%8
\end{eqnarray}  
Here, $s_l=5.14$ km/s and $s_t=3.04$ km/s are the sound velocities of longitudinal ($l$) and two transverse ($t$) modes, $\rho_M=5.31$ g/cm$^3$ is the crystal density, ${\cal D}=12$ eV is the deformation potential, and $h_{14}=1.2$ V/nm is the piezoelectric constant.
The contributions proportional to $\cos(4 \varphi_q)$ are not important in the isotropic model, as they are averaged out in Eq. (3). 
Further, we use the overlap integral, $I(q_z)=1/[(q_z b)^2+1]^3$, based on the Fang-Howard approximation, $\psi(z)=(2b^3)^{-1/2}ze^{-z/2b}$. 
The variational parameter $b$ is given by $b=[a_B/(33 n_e/8+ 12 n_d)]^{1/3}$, where $a_B =10$ nm is the Bohr radius and $n_d$ is the depletion charge density. 
We treat $b$ as an adjustable parameter, with a constraint $b < [8a_B/(33 n_e)]^{1/3} \approx 11$ nm, to get the closest fit between the theoretical and the experimental temperature dependence of the linear resistivity $\rho_0$. 
The electron-impurity scattering rate $\nim$ is extracted from $\rho_0$ at $T \rightarrow 0$, while the electron-phonon scattering rate $\nph$ is modeled with \req{eq.th.1} where we set $\te = T$. 
In \rfig{fig3}(a) we present measured (circles) and calculated (line) values of $\rho_0$ as a function of $T$. 
A good agreement is reached at $b=9.6$ nm, which we use in all further calculations. 
 
%%%%%%%%%%%%%%%%%%%%%%%%%%%%%%%%%%%%%%%%%%%%%%%%%
%%%%%%%%%%%%%%%%%%%%%%%%%%%%%%%%%%%%%%%%%%%%%%%%%
%

\begin{figure}[t]
\includegraphics{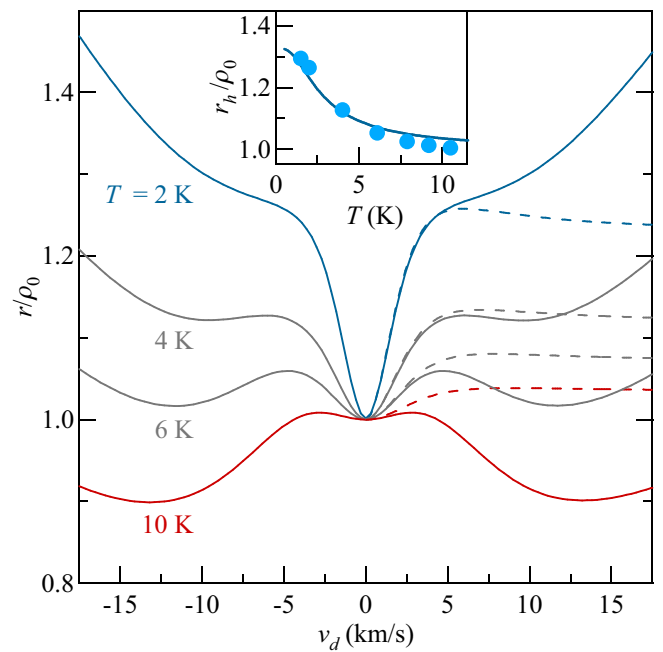}
\vspace{-0.1 in}
\caption{$r/\rho_0$ vs $\vd$ calculated for $T = 2,4,6$ and 10 K.
Dashed lines represent the degenerate electron gas approximation. 
Inset: Measured (circles) and calculated under degenerate approximation (line) $r_h/\rho_0$ vs $T$.
}
\vspace{-0.1 in}
\label{fig4}  
\end{figure}

In \rfig{fig3}(b) we present the calculated $\te$ averaged between the voltage probes, $\te=|x_1-x_2|^{-1} \int_{x_1}^{x_2} dx \te(x)$, at several $T$ from 2 K to 10 K, demonstrating that the electrons are strongly heated by $I$. 
The calculated values of $r/\rho_0$ are presented in \rfig{fig4} as a function of $\vd$ for several $T$. 
The results obtained in the approximation of degenerate electron gas and homogeneous heating are shown by dashed lines, demonstrating that such an approximation is reliable at small $v_d$ and low $T$. 
The $T$ dependence of $r_h/\rho_0$ obtained in this approximation is plotted in the inset of \rfig{fig4}, together with the experimental data. 
The characteristic features of the data shown in \rfig{fig2} are reproduced reasonably well. 
They include a rapid initial rise of $r$ with increasing $\vd$ at $T = 2$ K, a disappearance of this rise at $T \sim 10$ K, the nonmonotonic dependence of $r$ on $\vd$, and the asymmetry of $r$ with respect to the sign of $I$. 

To get a better quantitative agreement between theory and experiment, one needs to know the energy dependence of $\nim(\varepsilon)$, which was neglected in our calculations. 
Also, instead of the effective electron temperature approximation, one may apply more sophisticated approaches, e.g., based on the numerical Monte Carlo solution of the kinetic equation \citep{jacoboni:1983,haug:2007} taking into account a strongly inelastic interaction of electrons with optical phonons which becomes relevant at $\te \gtrsim 50$ K. 

%%%%%%%%%%%%%%%%%%%%%%%%%%%%%%%%%%%%%%%%%%%%%%%%%
%%%%%%%%%%%%%%%%%%%%%%%%%%%%%%%%%%%%%%%%%%%%%%%%

In summary, we have investigated nonlinear transport in a high-mobility 2DEG at electron 
drift velocities up to 20 km/s.
We identify two mechanisms of nonlinearity related to different energy scales determined by the current.
At small currents, the nonlinearity is caused by the heating of electrons above the Bloch-Gr\"uneisen temperature $\tbg$, which results in a rapid growth of the differential resistance if the lattice temperature is smaller than $\tbg$.
At higher currents, the nonlinearity reflects a breakdown of the state of a strongly degenerate electron gas, when the electron temperature becomes comparable to or exceeds the Fermi temperature so the differential resistance becomes sensitive to the energy dependence of the scattering time.
All the basic features observed are well explained by a spatially-inhomogeneous hot-electron model considering the interaction of electrons with impurities and acoustic phonons.
The observed effect represents a convenient technique for the detection of the Bloch-Gr\"uneisen regime and is promising for further studies of electron-phonon interactions in solids.
While the observation of a sizable resistance growth at low electron density requires high-mobility systems, we believe that the Bloch-Gr\"uneisen nonlinearity should also be observable in other systems of contemporary interest, including 2DEGs in MgZnO/ZnO heterostructures and graphene, in which $\tbg$ can be made high owing to the gate-induced enlargement of the electron density.
 
\begin{acknowledgements}
We thank L. Engel and I. Dmitriev for discussions, and S. Chakraborty for technical assistance.
The work at the University of Minnesota was funded by the NSF Grant No. DMR-1309578.
The work at the NHMFL/FSU was supported by DOE Grant No. DE-FG0205-ER46212.
The National High Magnetic Field Laboratory is supported by National Science Foundation Cooperative Agreement No. DMR-1157490 and the State of Florida.
This work was performed, in part, at the Center for Integrated Nanotechnologies, an Office of Science User Facility operated for the U.S. Department of Energy (DOE) Office of Science. 
Sandia National Laboratories is a multimission laboratory managed and operated by National Technology and Engineering Solutions of Sandia, LLC., a wholly owned subsidiary of Honeywell International, Inc., for the U.S. Department of Energy's National Nuclear Security Administration under Contract No. DE-NA-0003525.
\end{acknowledgements}

\end{document}